\documentclass[nofootinbib,aps,prl,preprint,superscriptaddress]{revtex4-1}
\usepackage{amsmath}
\usepackage{graphicx}
\usepackage[hidelinks]{hyperref} 
\usepackage{lineno}
\usepackage{bm}
\newcommand{\etal}{\textit{et al}.}

\begin{document}
\title{Enhanced thermal Hall conductivity below $1\,$Kelvin in the pyrochlore magnet Yb$_2$Ti$_2$O$_7$}
\author{Max Hirschberger}
\email{maximilian.hirschberger@riken.jp}
\altaffiliation{Current address: RIKEN Center for Emergent Matter Science (CEMS),Wako 351-0198, Japan}
\affiliation{Department of Physics, Princeton University, Princeton NJ 08540 USA}
\author{Peter Czajka}
\affiliation{Department of Physics, Princeton University, Princeton NJ 08540 USA}
\author{S. M. Koohpayeh}
\affiliation{Institute for Quantum Matter, Department of Physics \& Astronomy, Johns Hopkins University, Baltimore MD 21218 USA}
\author{Wudi Wang}
\affiliation{Department of Physics, Princeton University, Princeton NJ 08540 USA}
\author{N. Phuan Ong}
\affiliation{Department of Physics, Princeton University, Princeton NJ 08540 USA}
\date{\today}

\begin{abstract}
In this letter, we report the gigantic thermal Hall effect $\kappa_{xy}$ in the low-field correlated-paramagnetic state of the frustrated pyrochlore Yb$_2$Ti$_2$O$_7$. We observed a record magnitude for the thermal Hall angle in an insulator, $\left|\kappa_{xy}\right|/\kappa_{xx}\sim 2\,\%$. The signal onsets at $T\sim 3\,$K and is severely weakened around the transition to canted ferromagnetic order at $T_\text{CFM}=0.275\,$K. Besides the large $\kappa_{xy}>0$ of the fluctuating regime, a sign change towards negative $\kappa_{xy}$ occurs at the lowest temperatures and in moderate field, where sharp magnon excitations appear in the inelastic neutron scattering spectra. We analyze the magnon-Hall signal and its suppression with field semi-quantitatively. A contribution of phonon skew scattering to $\kappa_{xy}$ is ruled out by a comparison of $\kappa_{xx}$ for Tb-, Yb-, and Y-based rare earth pyrochlore titanates. These results represent the first report of non-vanishing $\kappa_{xy}$ measured in a dilution refrigerator ($T < 0.29\,$K).
\end{abstract}

\maketitle
\begin{figure}[htb]
  \begin{center}
		\includegraphics[width=0.7\linewidth]{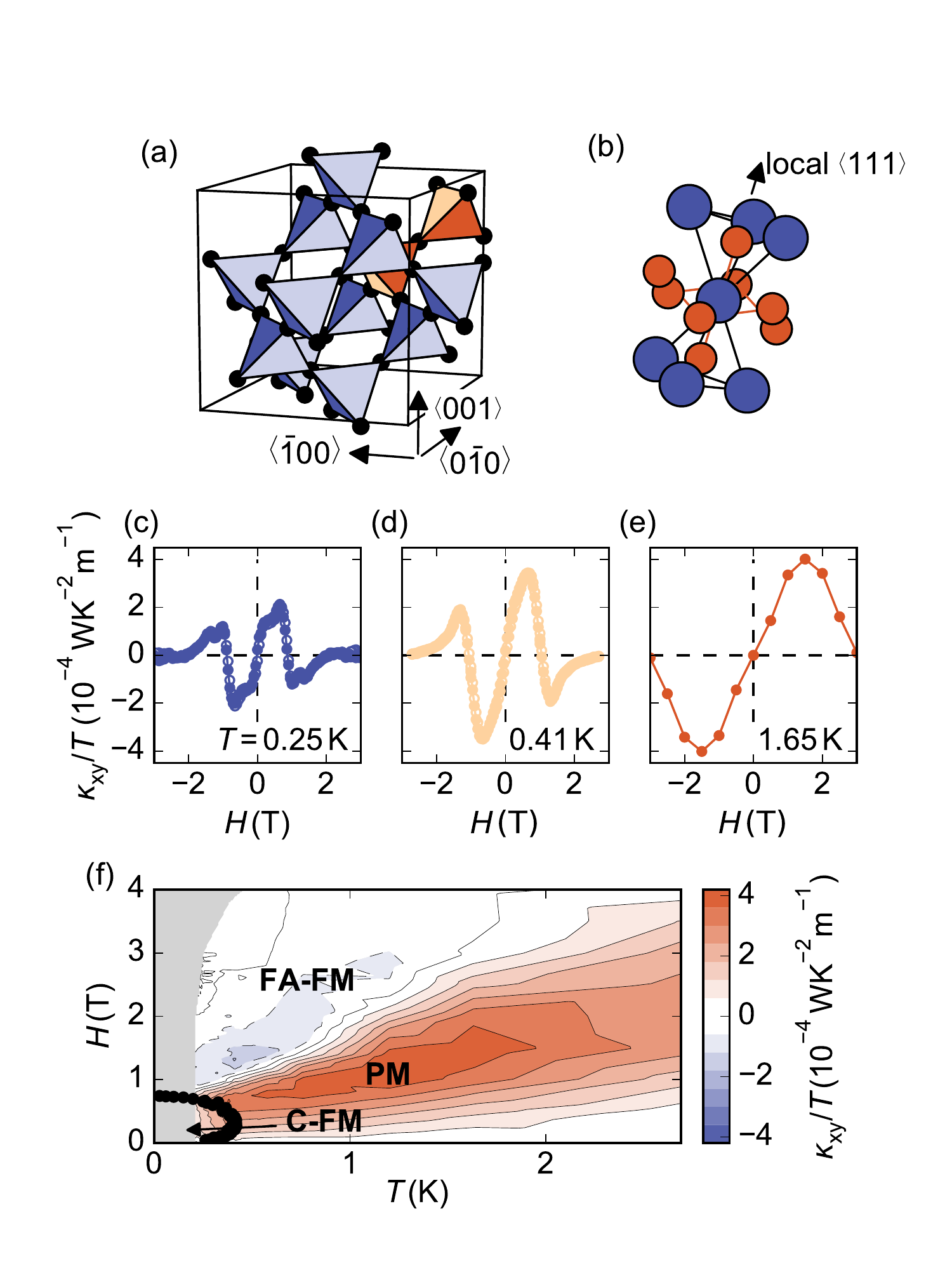}
    \caption[]{(color online). (a) Pyrochlore crystal structure of Yb$_2$Ti$_2$O$_7$. Only the rare-earth sublattice is shown (black dots), forming a network of corner-sharing tetrahedra. A pair of rare-earth tetrahedra is highlighted in red color. Crystallographic directions are indicated by arrows. (b) The crystal fields at each rare earth site (blue) are dominated by the surrounding oxygen ions (red). Ti$^{4+}$ ions not shown. Local $\left<111\right>$ direction for central rare earth ion is marked by an arrow. (c-e) Field dependence of $\kappa_{xy}/T$ at various $T$, with $\pmb{H} // \left<111\right>$, and heat current $\pmb{J}_Q // \left<110\right>$. (f) Contour plot of thermal Hall conductivity $\kappa_{xy}/T$. Black circles mark the phase transition between canted ferromagnetic (C-FM) order and the paramagnetic (PM) state (from Ref. \cite{Scheie2017}). FA-FM is the field-aligned ferromagnetic state.}
    \label{fig:fig1}
  \end{center}
\end{figure}

\begin{figure}[b]
  \begin{center}
		\includegraphics[width=0.85\linewidth]{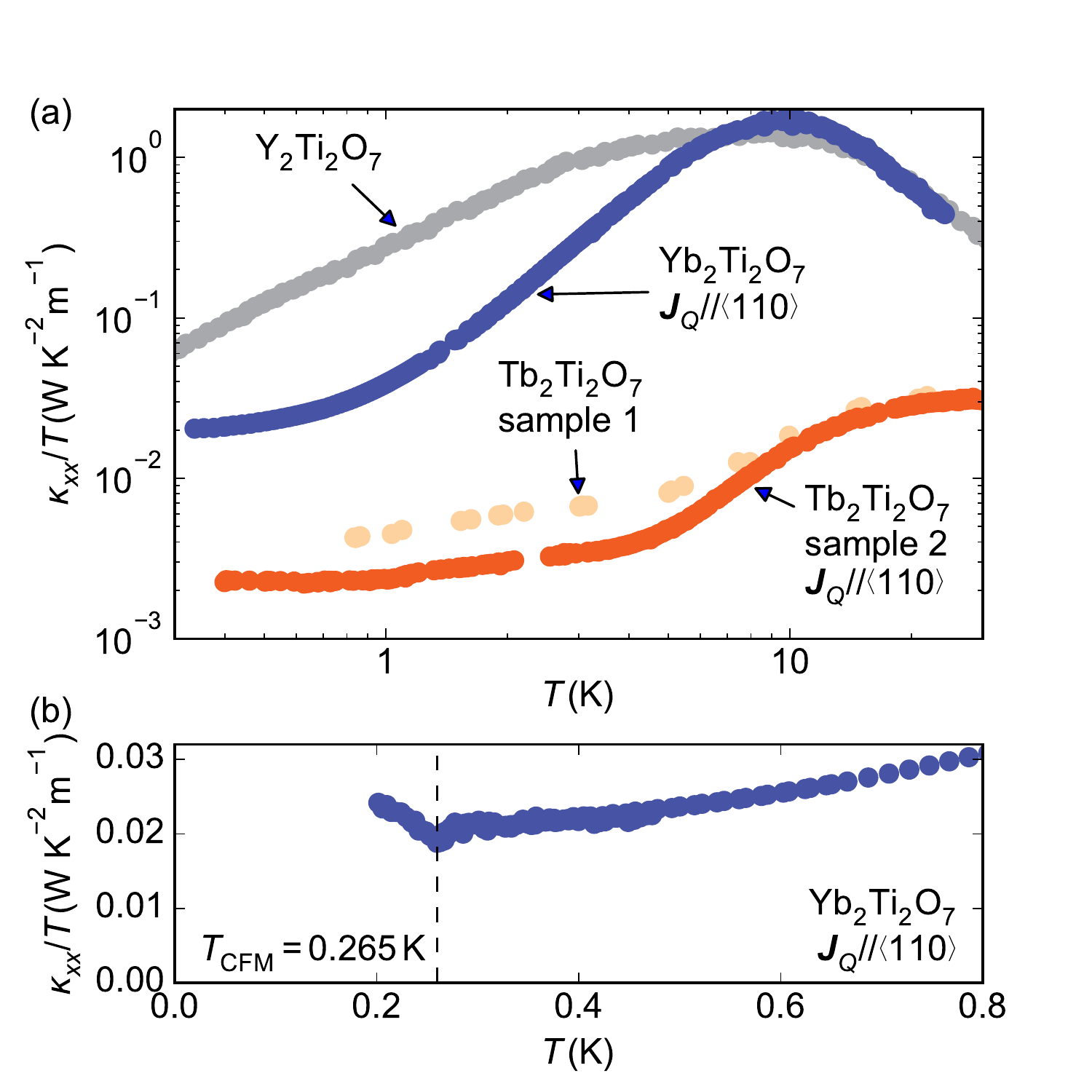}
    \caption[]{(color online). (a) Comparison of zero-field thermal conductivity $\kappa_{xx}$ for various $R_2$Ti$_2$O$_7$ compounds. At high $T\sim 20\,$K, the large maximum of $\kappa_{xx}$ in $R=$Yb indicates excellent sample quality. Phonon scattering at low $T$ is strongest in $R=$Tb (from Ref. \cite{HirschbergerScience2015}), and weakest in non-magnetic $R=$Y (from Ref. \cite{Li2013}). (b) Low-$T$ $\kappa_{xx}/T$ for $R=$Yb, showing a kink in the curve at the phase transition to long-range order (black dashed line).}
    \label{fig:fig2}
  \end{center}
\end{figure}

\begin{figure}[htb]
  \begin{center}
		\includegraphics[width=0.8\linewidth]{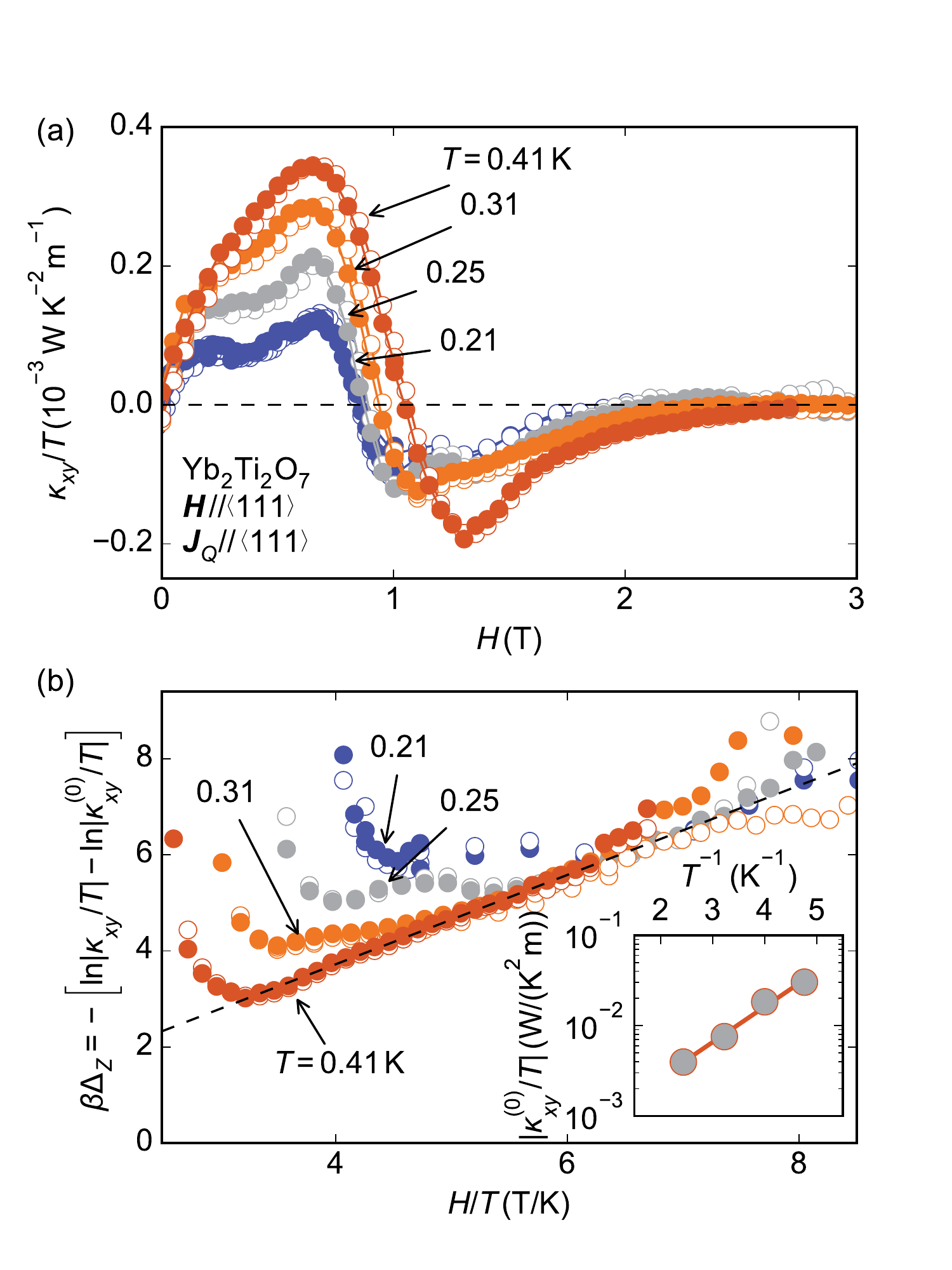}
    \caption[]{(color online). Scaling analysis of $\kappa_{xy}/T$ for Yb$_2$Ti$_2$O$_7$ in the field-aligned ferromagnetic state. (a) Raw data of the thermal Hall conductivity $\kappa_{xy}/T$; we focus on the negative signal observed for $\mu_0H > 1\,$T. (b) According to Eq. \ref{eq:kxy_scaling}, linear $\ln\left|\kappa_{xy}/T\right|\sim H/T$ is expected. A fit to the $T = 0.41\,$K data (dashed line) yields the effective $g$-factor (see text). The inset of (b) shows exponential $T$ dependence of the field-independent term $\kappa_{xy}^0/T$ from Eq. (2), from which the high-field anisotropy gap $\Delta_\text{A,HF}= 78\,\mu$eV was extracted by a fit (red line).}
    \label{fig:fig3}
  \end{center}
\end{figure}

When a condensed matter system is slightly nudged out of equilibrium by an applied heat current $J_{Q,i}$, quasiparticles such as electrons, magnetic excitations, and phonons drift along the resulting temperature gradient $\partial_i T$. The thermal transport tensor is defined as the linear response function relating these two quantities viz. $J_{Q,i} = \kappa_{ij}\partial_j T$ \cite{Ziman1964}. Broken time-reversal symmetry, e.g. through spontaneous ordering or an externally applied magnetic field $H$\cite{CommentField}, in principle allows for off-diagonal components of $\kappa_{ij}$: This is the thermal Hall or Righi-Leduc effect. Specializing to the case of insulators, $\kappa_{xy}\neq 0$, i.e. a transverse (Lorentz) force acting on charge-neutral quasiparticles, may seem surprising. Microscopic mechanisms for $\kappa_{xy}$ in insulators have therefore been considered from a theoretical perspective only over the past fifteen years \cite{Strohm2005}\cite{Katsura2010}\cite{Matsumoto2011}\cite{Murakami2017}.

In the absence of a classical Lorentz force, it is believed that an important and likely dominant contribution to $\kappa_{xy}$ in systems with non-interacting quasiparticles can be written as \cite{Katsura2010}\cite{Murakami2017}
\begin{equation}
\label{eq:kappaxy}
\kappa_{xy}=-\left(k_B^2 T\right)/V \sum_{n\pmb{k}}\Omega_{n\pmb{k}}^z c_2 \left(\rho_B \left(\epsilon_{n\pmb{k}} \right)\right) 
\end{equation}
where the summation is executed over all quasiparticle bands $\epsilon_{n\pmb{k}}$ with momentum $\pmb{k}$ and band index $n$. $\Omega_{n\pmb{k}}^z$ is the component of the Berry curvature vector parallel to the magnetization vector, $\rho_B$ is the appropriate distribution function (e.g. Bose-Einstein),  and $c_2\left(\rho_B\right)$ is a nonlinear, but monotonically decreasing function \cite{Murakami2017}. Eq. \ref{eq:kappaxy} for $\kappa_{xy}$ is explicitly independent of the mean free path of heat carriers, similar to the case of the intrinsic anomalous Hall conductivity in electron gases with non-trivial band structures \cite{Lee2004}\cite{Nagaosa2010}. 

Through Eq. \ref{eq:kappaxy}, it has become possible to quantitatively calculate the thermal Hall response of ferromagnets both in the long-range ordered state and above $T_C$ \cite{Onose2010}\cite{HirschbergerPRL2015}\cite{LeeHanLee2015}. This approach makes low-$T$ transport experiments powerful complements to spectroscopic techniques such as inelastic neutron scattering, particularly due to their sensitivity to the lowest lying excitations. In addition to a wealth of theoretical predictions of $\kappa_{xy}$ in various non-ferromagnetic long-ranged ordered magnets, the focus of recent theoretical and experimental research has moved towards frustrated magnets and spin liquids, their exotic elementary excitations, and associated $\kappa_{xy}$ \cite{Romhanyi2015}\cite{HirschbergerScience2015}\cite{Watanabe2016}\cite{Doki2018}\cite{Kasahara2018}. In this letter, we report on the gigantic $\kappa_{xy}$ in the fluctuating regime of a frustrated magnet, with high sensitivity to $H$ and the onset of long-range order at the lowest $T$. As previous work identified signatures of strongly interacting magnon states in the same region of the phase diagram where the Hall angle is large \cite{Thompson2017}\cite{Pecanha2017}, our experiment highlights the potential of strongly interacting or fractionalized excitations for generating gigantic $\kappa_{xy}$, possibly requiring a theoretical framework beyond Eq. \ref{eq:kappaxy}.

\textit{Target Material}. The frustrated pyrochlore titanate Yb$_2$Ti$_2$O$_7$ is a well-studied compound for which highly stoichiometric single crystals of excellent quality have recently been grown using a traveling-solvent float zoning technique \cite{Arpino2017}\cite{SI}. In the pyrochlore structure of corner-sharing tetrahedra, the crystal electric field (CEF) environment at the rare earth site (Fig. \ref{fig:fig1}) results in a simple level scheme for the $^2F_{7/2}$ Yb$^{3+}$ ion: the ground-state Kramers doublet is separated from all higher CEF states by an energy gap $\epsilon_1> 800\,$K \cite{Gaudet2015}. Then, the full exchange Hamiltonian $\mathcal{H}_{ex}$ is commonly expressed in terms of pseudospin-$1/2$ operators $S_i^z$ and $S_i^\pm$ defined with respect to the \textit{local} Ising axis (Fig. \ref{fig:fig1} (b)). 

As compared to the classical spin-ice systems Ho$_2$Ti$_2$O$_7$ and Dy$_2$Ti$_2$O$_7$ with very strong local-Ising anisotropy, with $\mathcal{H}_{ex} \approx \sum_{ij}J_{zz}S_i^zS_j^z$, and with macroscopic ground state degeneracy \cite{Anderson1956}\cite{Ramirez1999}\cite{Castelnovo2008}, fluctuations between up- and down-states are enhanced in Yb$_2$Ti$_2$O$_7$ due to the significant role of terms with $S_i^{+}$ and $S_j^{-}$ in $\mathcal{H}_{ex}$ \cite{Gingras2014}. These terms are sufficiently large to drive a transition to canted ferromagnetic order (C-FM) in powder samples \cite{Pecanha2017}\cite{Arpino2017} and stoichiometric single crystals \cite{Scheie2017}\cite{Arpino2017} at $T_\text{CFM} = 275\,$mK. Meanwhile, the paramagnetic state above $T_\text{CFM}$ shows hallmarks of strong correlations and fluctuations such as an anomalous increase of $T_\text{CFM}$ with $\left|H\right|>0$ \cite{Scheie2017}, pinch-points in diffuse neutron scattering \cite{Chang2012}, and strongly broadened, possibly gapless, inelastic neutron scattering spectra \cite{Thompson2017}\cite{Pecanha2017}. Significant spin-entropy is released up to $T = 3\,$K \cite{Arpino2017}, the characteristic energy scale of the dominant spin-spin interaction \cite{Hallas2017}.

\textit{Summary of experimental observations}. The key features of our thermal transport experiments, using a standard geometry with three semiconducting thermometers and a chip heater attached to a thin $\left<111\right>$ plate with $\pmb{J}_Q // \left<1\bar{1}0\right>$ \cite{SI}, are summarized in Fig. \ref{fig:fig1} (c-f): Very large $\kappa_{xy}/T>0$ (entropy factor removed) is observed at $T<3\,$K in the vicinity of zero field, until the signal is suppressed just below the phase transition to the C-FM state. Meanwhile, $\kappa_{xy}/T<0$ is a characteristic feature of the field-aligned ferromagnetic state (FA-FM, $H>1\,$T) at the lowest $T$. At very large fields, the thermal Hall signal is suppressed to zero, consistent with a large gap of magnetic excitations induced by the Zeeman effect.

\begin{figure}[tb]
  \begin{center}
		\includegraphics[width=1.0\linewidth]{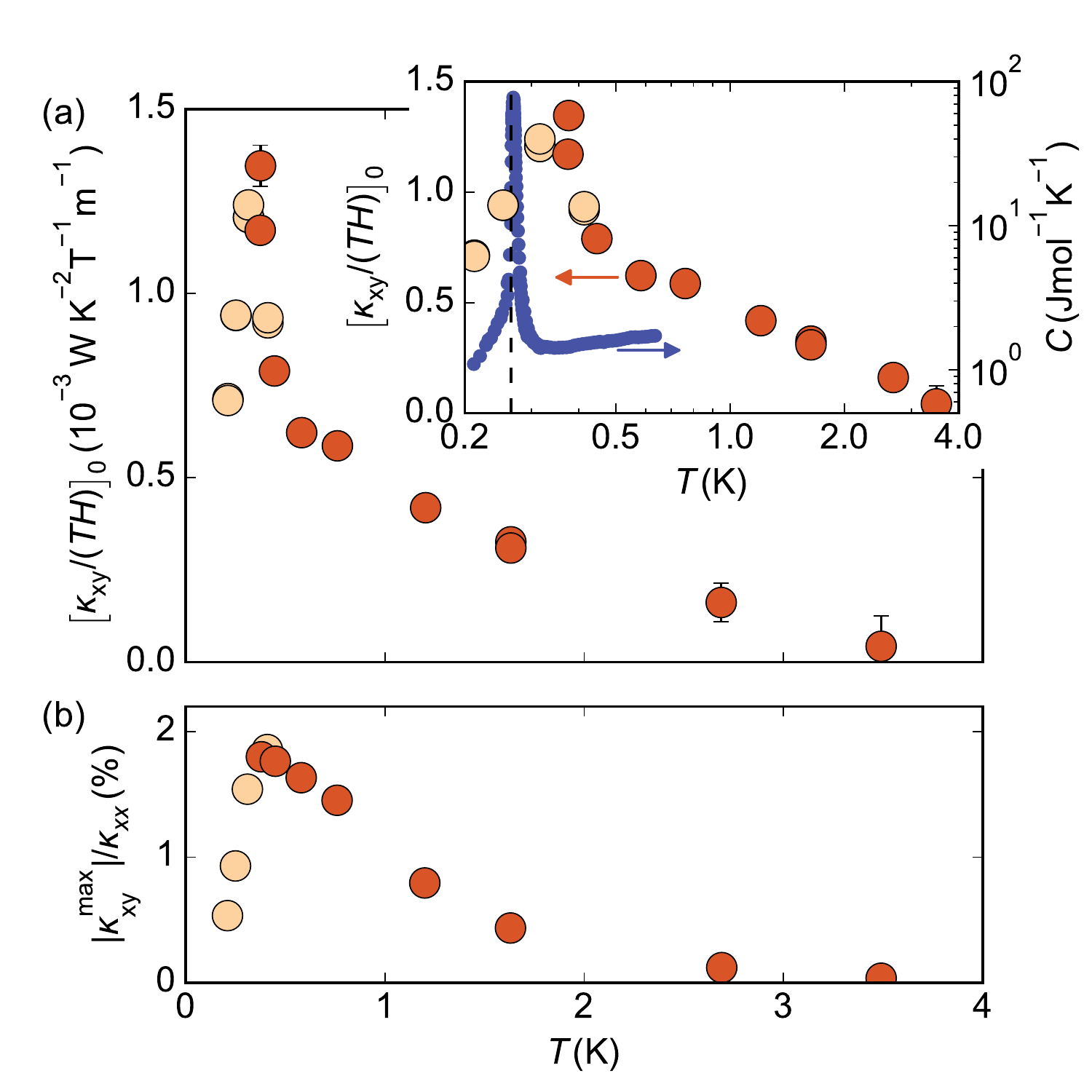}
    \caption[]{(color online). (a) The initial slope $\left[\kappa_{xy}/(TH)\right]_0$ of the thermal Hall conductivity in Yb$_2$Ti$_2$O$_7$.  Inset: On a logarithmic $T$-scale, the regime of largest $\left[\kappa_{xy}/(TH)\right]_0$ is clearly identified to be above the ordering transition (dashed black line). The ordering temperature $T_\text{CFM}$ was defined through the specific heat $c_P(T)$ from Ref. \cite{Arpino2017} (blue, right axis). $T_\text{CFM}$ of the thermal transport sample is $\sim 10\,$mK lower than in the case of $c_P(T)$ (c.f. main text). (b) Peak thermal Hall angle obtained by choosing the maximum value of each $\left|\kappa_{xy}\right|/\kappa_{xx}$ isotherm. Yellow symbols represent data recorded in a dilution refrigerator, while red data points were taken using a He-3 insert. }
    \label{fig:fig4}
  \end{center}
\end{figure}

Our sample was taken from the same batch as the one used for specific heat $c_P(T)$ measurements in Ref. \cite{Scheie2017}. At $T_\text{CFM}$, where a sharp spike was observed in $c_P(T)$ (Ref. \cite{Arpino2017}\cite{Scheie2017} and Fig. \ref{fig:fig4}, inset), we found a kink in $\kappa_{xx}(T)$ in zero field (Fig. \ref{fig:fig2} (b)). However, we also found that our $T_\text{CFM} = 0.265\,$K was slightly reduced as compared to the previous report ($T_\text{CFM}^\text{Arpino} = 0.275\,$K)\cite{Arpino2017}\cite{SI}. This may be associated with a mild deterioration of the sample quality during the preparation process \cite{SI}. 

\textit{Discussion of phonon skew scattering}. Recently, $\kappa_{xy}$ in rare earth pyrochlore titanates has been discussed not only from the point of view of the intrinsic (Berry phase) theory of Eq. \ref{eq:kappaxy}, but also using an extrinsic scattering mechanism (phonon skew scattering) \cite{Strohm2005}\cite{Onose2019}. In the present work, we demonstrate that $\kappa_{xy}/T$ in Yb$_2$Ti$_2$O$_7$ exhibits highly nonlinear behavior as a function of $H$ and $T$, qualitatively at odds with the extrinsic scenario. We observed vanishing $\kappa_{xy}$ when $M$ approaches saturation, suppression of $\kappa_{xy}/T$ around the ordering temperature, and even a sign change at intermediate $H$ and low $T$. 

A comparison of $\kappa_{xx}(T)$ for three rare earth titanates $R_2$Ti$_2$O$_7$ ($R$: rare earth) sheds further light on the phonon scattering behavior: (i) Non-magnetic $R=$Y has negligible $\kappa_{xy}$ \cite{HirschbergerScience2015} and has the largest $\kappa_{xx}$ at low $T$. This corresponds to a large phonon mean-free path in the absence of resonant scattering from magnetic rare earth sites \cite{Li2013}. (ii) $\kappa_{xx}$ is strongly suppressed in $R=$Tb with non-Kramers ion Tb$^{3+}$ due to coupling of the ground state quasi-doublet to lattice vibrations. Experimentally, the observation of a hybrid magnetoelastic excitation in inelastic neutron scattering \cite{Fennell2014} is perhaps the clearest evidence of strong spin-lattice coupling in $R=$Tb, resulting in a very short phonon mean free path $\sim 1\,\mu$m \cite{Li2013}. (iii) In contrast, Yb$^{3+}$ is a Kramers ion, i.e. spin-phonon coupling cannot lift the degeneracy of the ground state doublet and we expect phonon skew scattering to be suppressed. Although $\kappa_{xx}$ of $R=$Yb lies in an intermediate regime (Fig. \ref{fig:fig2} (a)), its $\kappa_{xy}/T$ and thermal Hall angle $\left|\kappa_{xy}\right|/\kappa_{xx}$ are even larger than those of $R=$Tb \cite{SI}. We conclude that the thermal Hall signal reported in this manuscript is inconsistent with the phonon skew scattering mechanism \cite{Strohm2005}.

\textit{Field-aligned ferromagnetic state.} In the FA-FM state, the presence of sharp magnon quasiparticle modes has been established convincingly in inelastic neutron scattering and THz experiments \cite{Ross2011}\cite{Pan2014}\cite{Thompson2017}. This is the regime where $\kappa_{xy}<0$ was observed in our experiment at $1\,\text{T}<H<3\,\text{T}$ (Fig. \ref{fig:fig3}). Using the experimental magnon spectra, repeated attempts have been undertaken to determine the coupling constants in $\mathcal{H}_{ex}$. For example, the seminal work by Ross \etal{} reported negligible dipolar interactions and dominant spin-ice (Ising-like) correlations $\sim J_{zz}S_i^zS_j^z$ in $\mathcal{H}_{ex}$ \cite{Ross2011}. Meanwhile, fluctuation terms $\sim S_i^{+}S_j^{-}$, $\sim S_i^{+}S_j^{+}$, $\sim S_i^{z}S_j^{+}$, etc. were also found to be significant \cite{Ross2011}. A successive experimental study, exploring a wide range of reciprocal space, reported Hamiltonian parameters at odds with the initial work \cite{Thompson2017}; ambiguities thus remain concerning the reliability of such fitting schemes.

Our thermal transport study provides a highly desirable, independent check on these neutron scattering experiments, clarifying firstly that there are no additional lower-energy magnon bands with finite Berry curvature at $H>1\,$T. Secondly, although this is beyond the scope of the present paper, the $\kappa_{xy}$ data can be modeled quantitatively when assuming a given set of Hamiltonian parameters. Thirdly, some key parameters describing the magnon spectrum in the FA-FM state can be extracted by treating $\kappa_{xy}$ phenomenologically. When a Zeeman gap is opened in $H\neq 0$, the $T$-dependence of $\kappa_{xy}$ is dominated by changes in the number density of heat carrying excitations and we crudely simplify Eq. \ref{eq:kappaxy} to
\begin{align}
\label{eq:kxy_scaling}
\kappa_{xy}/T&=\left(\bar{\kappa}_{xy}/T\right)\cdot\exp⁡\left(-\beta\left(\Delta_\text{A,HF}+\Delta_\text{Z,HF}\right)\right)\notag\\
&=\left(\kappa_{xy}^{(0)}/T\right)\cdot\exp\left(-\beta\Delta_\text{Z,HF}\right)
\end{align}
The total spin gap is a combination of $H$-dependent Zeeman gap $\Delta_\text{Z,HF} = (g_\text{eff}/2)\mu_B H$ and $H$-independent anisotropy gap $\Delta_\text{A,HF}$. Here, $\mu_B$ is the Bohr magneton and $\beta=(k_BT)^{-1}$ with Boltzmann constant $k_B$. We have introduced an averaged, effective $g$-factor defined for pseudospin-$1/2$, but the reader should be aware that neutron scattering experiments have reported significant anisotropy of $g$ in Yb$_2$Ti$_2$O$_7$ \cite{Thompson2017}\cite{Ross2011}. The reduced $\left(\kappa_{xy}^{(0)}/T\right)$ may contain additional, subdominant power-law $T$-dependence \cite{Onose2010}.

We test Eq. \ref{eq:kxy_scaling} using a scaling analysis in the FA-FM state as shown in Fig. \ref{fig:fig3} (b). All the curves of $\ln\left|\kappa_{xy}/T\right|$ are found to collapse onto a universal line as a function of $\beta\Delta_\text{Z,HF} \sim H/T$, if a $H$-independent term $\sim \beta\Delta_\text{A,HF}$ is subtracted. This yields $g_\text{eff} = 2.8$. Fitting the $H$-independent term $\ln\left|\kappa_{xy}^{(0)}/T\right|$ as a function of $T$, we also estimate the anisotropy gap in the FA-FM state: $\Delta_\text{A,HF} = 78\,\mu$eV (Fig. \ref{fig:fig3} (c)). Note that $\Delta_\text{A,HF}$ in the FA-FM state is distinct from excitation gap of the C-FM state at low $H$. 

\textit{Thermal Hall effect in the correlated paramagnetic state.} 
As alluded to earlier in the text, the fluctuating correlated-paramagnetic state of Yb$_2$Ti$_2$O$_7$ hosts a very large \textit{positive} $\kappa_{xy}$ signal just above $T_\text{CFM}$. We study the linear response in this regime through the low-$H$ slope of the thermal Hall conductivity $\left[\kappa_{xy}/(TH)\right]_0 = \left[\kappa_{xy}/(TH)\right]_{H\rightarrow0}$ (Fig. \ref{fig:fig4}). The onset of finite $\left[\kappa_{xy}/(TH)\right]_0$ occurs just below $T = 3\,$K, in good agreement with a hump in $c_P(T)$ around the same $T$ \cite{Hodges2002}. As $T$ is lowered further, $\left[\kappa_{xy} /(TH)\right]_0$ increases sharply below $0.5\,$K, reaches a maximum around $0.35\,$K, and then finally starts to decrease as $T_\text{CFM}$ is approached from above and transgressed (Fig. \ref{fig:fig4}). We compare with $c_P(T)$ for a sample from the same batch (inset of Fig. \ref{fig:fig4}, adapted from Ref. \cite{Arpino2017}). The presence of a maximum in $\left[\kappa_{xy}/(TH)\right]_0$ at $T = 0.4\,$K strongly suggests that the spin gap of the PM state should be smaller than $40\,\mu$eV.

A regime of enhanced $\left[\kappa_{xy} /(TH)\right]_0$ just above $T_\text{CFM}$ as reported here was foreshadowed in a combined study of elastic neutron scattering and bulk techniques by Scheie \etal{} \cite{Scheie2017}. These authors reported that small $\pmb{H} //\left<111\right>$ can suppress fluctuations in the correlated-paramagnetic state, increasing $T_\text{CFM}$ and resulting in an unconventional lobe-shaped phase diagram (Fig. \ref{fig:fig1} (f), (black symbols) \cite{Scheie2017}. In our data, the largest $\left[\kappa_{xy}/(TH)\right]_0$ is observed in the fluctuation region identified by Ref. \cite{Scheie2017}. In fact, the positive thermal Hall signal takes on record values as compared to $\kappa_{xx}$. We report the maximum value of the Hall angle $\left|\kappa_{xy}^{max}\right|/\kappa_{xx}$ as a function of $T$ in Fig. \ref{fig:fig4} (b). The peak magnitude of nearly $2\,\%$ is about three times larger than any previously reported signal \cite{Ideue2017}.

\textit{Discussion and conclusion.} Previously, some attempts have been made in systems with long-range order to model the thermal Hall signal in the correlated-paramagnetic regime using Schwinger-boson linearization of the Hamiltonian and Eq. \ref{eq:kappaxy}. In the layered Kagome ferromagnet Cu(1,3-bdc), sizable $\kappa_{xy}$ observed above the onset of long-range order \cite{HirschbergerPRL2015} was explained in this framework \cite{LeeHanLee2015}. A weaker, $H$-linear signal appears in frustrated Kagome systems with antiferromagnetic nearest-neighbor exchange and strongly suppressed, yet non-zero, $T_N$ \cite{Watanabe2016}\cite{Doki2018}. 

For Yb$_2$Ti$_2$O$_7$ we speculate that the approximation of non-interacting particles underlying Eq. \ref{eq:kappaxy} may break down completely in the correlated-paramagnetic regime, especially around $T_\text{CFM}$. This hypothesis is supported by the experimental observation of a very broad inelastic scattering continuum in Yb$_2$Ti$_2$O$_7$ at low $H$, even below $T_\text{CFM}$ \cite{Thompson2017}\cite{Pecanha2017}\cite{OnodaPC}, which may be connected to strongly interacting two-magnon states \cite{Thompson2017}. In this scenario, a quantitative comparison of our low-$H$ $\kappa_{xy}$ data to theory requires a generalization of the standard framework of Refs. \cite{Matsumoto2011}\cite{Murakami2017} (Eq. \ref{eq:kappaxy}) to the case of strongly interacting magnon states \cite{Rau2019}. Alternatively, Eq. \ref{eq:kappaxy} may retain its validity if the broad neutron scattering spectrum is related to quasiparticle fractionalization \cite{Rau2019}\cite{Han2012}\cite{Banerjee2016}. In either case, the experiment demonstrates that gigantic thermal Hall responses can be generated from magnetic excitations in the absence of long-range order. We expect that our work will further the ongoing search \cite{Ideue2017} for giant $\kappa_{xy}$ and related spin-Hall and spin-Nernst effects in the solid state.

\textit{Acknowledgments.} We thank R.J. Cava and J. Krizan for initial attempts at float zoning crystal growth for Yb$_2$Ti$_2$O$_7$. We also acknowledge stimulating discussions with S. Onoda. C. Pfleiderer and S. S\"aubert kindly made their low-$T$ magnetization data from Ref. \cite{Scheie2017} available to us. The research at Princeton was supported by the Department of Energy (DE-SC0017863), the Gordon and Betty Moore Foundation's EPiQS initiative through Grants GBMF4539, and the U.S. National Science Foundation (Grant DMR 1420541). The work at the Institute for Quantum Matter, an Energy Frontier Research Center, was supported by the U.S. Department of Energy, Office of Science, Office of Basic Energy Sciences under Award Number DE-SC0019331.

\end{document}